\documentclass[pra, 12 pt]{revtex4}
\usepackage[english]{babel}
\usepackage{amsmath}
\usepackage{epsfig}

\begin{document}
\title{Generalized Master Equation with Two Times: Diffusion in External Field}
\author{S.A. Trigger}
\address{Joint\ Institute\, for\, High\,
Temperatures, Russian\, Academy\, of\, Sciences, 13/19, Izhorskaia
Str., Moscow\, 127412, Russia;\\
email:\,strig@gmx.net}

\begin{abstract}
The generalized master equation with two times, introduced in
[1,2], applies to the problem of diffusion in an time-dependent
(in general inhomogeneous) external field. We consider the case of
the quasi Fokker-Planck approximation, when the probability
transition function for diffusion (PTD-function) does not possess
a long tail in coordinate space and can be expanded as the
function of instantaneous displacements. The more complicated case
of the long tails in PTD will be discussed separately.

\end{abstract}.

\maketitle

\section{Introduction}

The models of continuous time random walks (CTRW) [3], when some
objects can jump from one point to another in inhomogeneous (in
general) media and stay during some time in these points before
the next usually stochastic jump, are important for the solution
of many physical, chemical and biological problems. Recently these
models have been applied also in economics and social sciences
(see, e.g., [4-6]). Usually the stochastic motion of the particles
leads to a second moment of the density distribution that is
linear in time $<r^2(t)>\sim t$. Such type of the diffusion
processes play a crucial role in plasmas, including dusty plasma
[7], nuclear physics [8], neutral systems in various phases [9]
and in many other problems. However, in many systems the deviation
from the linear in time dependence of the mean square displacement
have been experimentally observed, in particular, under
essentially non-equilibrium conditions or for some disordered
systems. The average square separation of a pair of particles
passively moving in a turbulent flow grows, according to
Richardson's law, with the third power of time [10]. For diffusion
typical for glasses and related complex systems [11] the observed
time dependence is slower than linear. These two types of
anomalous diffusion obviously are characterized as superdiffusion
and subdiffusion.

The generalized master equation for the density evolution, which
describes the various cases of normal and anomalous diffusion has
been formulated in [1,2] by introduction of the specific kernel
function (PTD) depending on two times $W({\bf r, \bf r'}, \tau,
t-\tau)$, which connects in a linear way the density distributions
of the stochastic objects (or particles) $f$ for the points ${\bf
r'}$ at moment $\tau$ and ${\bf r}$ at moment $t$. The approach
suggested in [1,2] clearly demonstrates the relation between the
integral approach and the fractional differentiation method [12]
and permits to extend (in comparison with the fractional
differentiation method) the class of sub- and superdiffusion
processes, which can be successfuly described. On this basis in
[2] the different examples of superdiffusive and subdiffusive
processes were considered for the various kernels $W$ and the
mean-squared displacements have been calculated. Recently the idea
of the generalized master equation with two times [1,2] for
diffusion in coordinate space has been recently used in [13] for
the calculation of average displacements in the case of a
time-dependent homogeneous external field.

This paper is motivated by the necessity to describe in more
detail the influence of time- and space dependent external fields
on the continuous-time random walks. The equation formulated in
[1,2] is appropriate for this purpose and offers the opportunity
for consideration of CTRW for both cases: long-tail space behavior
of the PTD function, as well as for the fast decay of PTD function
in coordinate space, when the Fokker-Planck type expansion is
applicable. For simplicity in this paper we consider only the last
case.

\section{Generalized Master equation}

Let us start from the generalized master equation with two times
[1,2]:
\begin{equation}
f({\bf r},t) = f({\bf r},t=0)+ \int_0^t d\tau \int d{\bf r'}
\left\{W ({\bf r, r'},\tau, t-\tau) f({\bf r',\tau}) - W ({\bf r',
r},\tau, t-\tau) f({\bf r},\tau) \right\}. \label{F1}
\end{equation}
Equation (\ref{F1}) can be represented in an equivalent form, more
similar to the structure of the Fokker-Planck equation, where the
initial condition is absent:
\begin{equation}
\frac{\partial f({\bf r},t)}{\partial t} = \frac{d}{dt} \int_0^t
d\tau \int d{\bf r'} \left\{W ({\bf r, r'},\tau, t-\tau) f({\bf
r',\tau}) - W ({\bf r', r},\tau, t-\tau) f({\bf r},\tau) \right\}.
\label{F2}
\end{equation}
or
\begin{equation}
\frac{\partial f({\bf r},t)}{\partial t} = \int_0^t d\tau \int
d{\bf r'} \left\{P ({\bf r, r'},\tau, t-\tau) f({\bf r',\tau}) - P
({\bf r', r},\tau, t-\tau) f({\bf r},\tau) \right\}, \label{F3}
\end{equation}
where the function $P ({\bf r, r'},\tau, t-\tau)$ is given by:
\begin{equation}
P ({\bf r, r'},\tau, t-\tau)\equiv 2W ({\bf r', r},\tau,
t-\tau)\delta (t-\tau)+ \frac{\partial}{\partial t} W ({\bf r',
r},\tau, t-\tau)\label{F4}
\end{equation}
The argument $t-\tau$ describes the retardation effects, which can
be connected in the particular case of multiplicative PTD function
$W ({\bf r, r'},\tau, t-\tau)\equiv W ({\bf r, r'},\tau)
\chi(t-\tau)$ with, for example, the probability for the particles
to stay during some time at the fixed coordinate before transition
to the next point. An equation with retardation, where $W$
function depends only on one time argument $t-\tau$ has been
suggested first in [14] and applied in [15] to the case of the
multiplicative representation of the PTD function. In general $W$
is not a multiplicative function in the sense mentioned above and,
what is more important, is a function of two times $t$ and
$t-\tau$ [1]. It is necessary to mention that the closed form of
the equation for the density distribution is an approximation. In
some cases the exact solution for density distribution can be
found (see e.g. [16],[17]), when the closed equation for density
distribution does not exist or gives an approximate result.
Nevertheless, in many practical situations Eq.~(\ref{F1}) or
(\ref{F4}) are sufficiently exact and permit to describe various
experimental data.

Let us consider the nature of appearance of the two time arguments
in the generalized master equation Eq.~(\ref{F1}) in the case of a
time-dependent external force ${\bf F}({\bf r},t)$. To simplify
the consideration we can investigate the case of fast decay of the
kernel $W ({\bf r, r'},\tau, t-\tau)\equiv W ({\bf u, r},\tau,
t-\tau)$ as a function of ${\bf u=r-r'}$, when an expansion in the
spirit of Fokker-Planck can be applied. In this case
Eq.~(\ref{F1}) takes the form [1,2]:
\begin{eqnarray}
f({\bf r},t) = f({\bf r},t=0)+ \int_0^t d\tau {\partial \over
{\partial r_\alpha}} \left[A_\alpha ({\bf r},\tau, t-\tau) f({\bf
r},\tau) + {\partial \over {\partial r_\beta}} \left(
B_{\alpha\beta}({\bf r},\tau, t-\tau) f({\bf r},\tau)
\right)\right], \label{F5}
\end{eqnarray}
where the functions $A_\alpha ({\bf r},\tau, t-\tau)$ and
$B_{\alpha\beta}({\bf r},\tau, t-\tau) f_g({\bf r},\tau)$ are the
functionals of the PTD function (the indexes are equal
$\alpha,\beta=x_s$ in s-dimensional coordinate space):
\begin{eqnarray}
A_\alpha({\bf r},\tau, t-\tau) = \int d^s u u_\alpha W({\bf u,
r},\tau, t-\tau) \label{F6}
\end{eqnarray}
and
\begin{eqnarray}
B_{\alpha\beta}({\bf r},\tau, t-\tau)= \frac{1}{2}\int d^s u \,
u_\alpha u_\beta W({\bf u, r},\tau, t-\tau). \label{F7}
\end{eqnarray}
Eq.~(\ref{F5}) can be rewritten naturally in the form,
corresponding to Eq.~(\ref{F2}), but now for the Fokker-Planck
type approximation:
\begin{eqnarray}
\frac{\partial f({\bf r},t)}{\partial t} = \frac{d}{dt} \int_0^t
d\tau {\partial \over {\partial r_\alpha}} \left[A_\alpha ({\bf
r},\tau, t-\tau) f({\bf r},\tau) + {\partial \over {\partial
r_\beta}} \left( B_{\alpha\beta}({\bf r},\tau, t-\tau) f({\bf
r},\tau) \right)\right], \label{F2a}
\end{eqnarray}
We suggest, that the PTD function is independent of $f({\bf
r},t)$, therefore the problem is linear.

\section{Influence of the external fields}

One of the main sources of inhomogeneity is an external field,
which also provides the prescribed dependence of the PTD function
on $\tau$. Other wards we can suggest, in the considered
particular case, that the dependence of $W({\bf u, r},\tau,
t-\tau)$ on the arguments ${\bf r},\tau$ is connected with a
functional dependence on an external field:
\begin{equation}
W({\bf u, r},\tau, t-\tau)=W({\bf u}, t-\tau; {\bf F}({\bf
r},\tau)).\label{F8}
\end{equation}
If the external fields are absent the PTD function is a function
of the modulus ${\bf u}\equiv u$, which means that $A_\alpha=0$
and $B=\delta_{\alpha\beta}B_0(t-\tau)$ with:
\begin{equation}
B_0(t-\tau)= \frac{1}{2s}\int d^s u \, u^2
W_0(u,t-\tau).\label{F7a}
\end{equation}

For relatively weak external fields the functional (\ref{F8}) can
be linearized in the external field
\begin{equation}
W({\bf u}, t-\tau; {\bf F}({\bf r},\tau))=W_0(u, t-\tau)+ W_1(u,
t-\tau)({\bf u} \cdot {\bf F}({\bf r},\tau)). \label{F9}
\end{equation}
The functions $W_0(u, t-\tau)$ and $W_1(u, t-\tau)$ are equal to
$W({\bf u}, t-\tau; {\bf F}=0)$ and the functional derivative
$\delta W({\bf u}, t-\tau; {\bf F}({\bf r},\tau))/\delta({\bf u}
\cdot {\bf F}({\bf r},\tau))_{|\textbf{F}=0}$ respectively. Then
the functions $A_\alpha$ and $B_{\alpha\beta}$ take the form
\begin{eqnarray}
A_\alpha({\bf r},\tau, t-\tau) =\frac{1}{s}{\bf F}_\alpha ({\bf
r},\tau)\int d^s u u^2 W_1(u, t-\tau)\equiv {\bf F}_\alpha ({\bf
r},\tau)L(t-\tau),\label{F10}
\end{eqnarray}
where $L(t-\tau)$ is given by
\begin{eqnarray}
L(t-\tau) =\frac{1}{s}\int d^s u u^2 W_1(u, t-\tau).\label{F10a}
\end{eqnarray}
and
\begin{eqnarray}
B_{\alpha\beta}({\bf r},\tau, t-\tau)=
\delta_{\alpha\beta}B_0(t-\tau). \label{F11}
\end{eqnarray}
The generalized diffusion equation Eq.~(\ref{F2a}) takes the form
\begin{eqnarray}
\frac{\partial f({\bf r},t)}{\partial t} = \frac{d}{dt} \int_0^t
d\tau \left[L(t-\tau)\nabla({\bf F} ({\bf r},\tau) f({\bf
r},\tau)) + B_0(t-\tau){\Delta} f({\bf r},\tau)\right],
\label{F2b}
\end{eqnarray}
In general this equation contains two different functions $B_0$
and $L$ depending on the argument $t-\tau$. If the functional
$W({\bf u}, t-\tau; {\bf F}({\bf r},\tau))$ is multiplicative,
namely, $W({\bf u}, t-\tau; {\bf F}({\bf r},\tau))=\tilde W({\bf
u}; {\bf F}({\bf r},\tau))\chi(t-\tau)$ Eq.~(\ref{F2b}) can be
simplified:
\begin{eqnarray}
\frac{\partial f({\bf r},t)}{\partial t} = \frac{d}{dt} \int_0^t
d\tau \chi(t-\tau)\left[D {\Delta} f({\bf r},\tau)-b\nabla({\bf F}
({\bf r},\tau) f({\bf r},\tau))\right], \label{F2c}
\end{eqnarray}
Here $b$ and $D$ are the constants, determined by the relations:
\begin{eqnarray}
b =-\frac{1}{s}\int d^s u u^2 \tilde W_1(u)\label{F10b}
\end{eqnarray}
with $\tilde W_1(u)=\delta \tilde W({\bf u}; {\bf F}({\bf
r},\tau))/\delta({\bf u} \cdot {\bf F}({\bf
r},\tau))_{|\textbf{F}=0}$ and
\begin{eqnarray}
D =\frac{1}{2s}\int d^s u u^2 \tilde W_0(u).\label{F10a}
\end{eqnarray}
The physical sense of the multiplicative structure of the
functional $W$ is that independence of the time delay of the
random walkers is independent of the external field. The
dimensionless function $\chi(t)$ in this simple case is connected
with the hopping-distribution function $\psi(t)=\lambda
\psi^\ast(\lambda t)$ introduced in the master equation by Scher
and Montroll [15]. The value $\lambda\equiv 1/\tau_0$ is the
characteristic waiting time for the hopping-distribution. Laplace
transformations of these functions $\chi(z)$ and $\psi^\ast(z)$
relate them as follows
\begin{eqnarray}
\chi(z)=\frac{\psi^\ast(z)}{1-\psi^\ast(z)}.\label{F12}
\end{eqnarray}
For the exponential hopping-time distribution $\psi(t)=\lambda
exp(-\lambda t)$, where $\lambda\equiv1/\tau_0$ ($\tau_0$ is the
characteristic waiting time) we have $\psi^\ast(z)=1/(1+z)$,
$\chi(z)=1/z$ and $\chi(t)\equiv\chi(\lambda t)=1$. In this case
Eq.~(\ref{F2c}) is reduced to the usual diffusion equation in an
external field with the diffusion coefficient $D$ and mobility
$b$:
\begin{eqnarray}
\frac{\partial f({\bf r},t)}{\partial t} = D {\Delta} f({\bf
r},t)-b\nabla\left({\bf F} ({\bf r},t) f({\bf r},t)\right).
\label{F13}
\end{eqnarray}

\section{Conclusions}

We show that the generalized master equation with two times, which
have been introduced in [1,2] can describe the influence of
inhomogeneous and time-dependent external fields on the diffusion
processes. Linearization of the general master equation in the
external field leads to essential simplifications. In this case
the diffusion processes depend, in general, on two different
functions of time, which describe retardation due to the finite
time of occupation and transferring particles in space in the
presence of the external field. Relations with simpler models are
established. The consideration in the present paper give the
opportunity to consider a wide class of the problems of normal and
anomalous transport in external fields on the basis of generalized
master equations with two times.

\section*{Acknowledgment}
The authors are thankful to A.M. Ignatov, P.P.J.M. Schram and
Yu.P. Vlasov for valuable discussions of the problems, reflected
in this paper. This work has been supported by The Netherlands
Organization for Scientific Research (NWO) and the Russian
Foundation for Basic Research.

\end{document}